\documentclass[12pt]{article}
\date{}
\usepackage{latexsym}
\usepackage{epsfig}
\begin{document}

\rightline{CU-TP-937}
\vskip30pt

\begin{center}

{\large\bf PARTON SATURATION AT SMALL\ x\ AND IN LARGE NUCLEI}
\vskip20pt
{A.H. Mueller\footnote{This research is sponsored in part by the Department of Energy, Grant\\
DE-FG02-94ER-40819.}\\ Department of Physics, Columbia University\\
New York, New York 10027}
\end{center}

\vskip30pt
\begin{abstract}
Quark and gluon distributions in the light-cone wavefunction of a high energy hadron or nucleus
are calculated in the saturation regime.  One loop calculations are performed explicitly using the
equivalence between the parton distribution in the light-cone wavefunction  and the production
distribution of that parton in a current-nucleon (nucleus) scattering.  We argue that, except for
some overall numerical factors, the Weizs\"acker-Williams wavefunction correctly gives the
physics of the gluon distribution in a light-cone wavefunction.
\end{abstract}

\pagenumbering{arabic}

\section{Introduction}
\indent The idea of parton saturation\cite{Gri} in QCD is at the heart of the interest in small-x  hadron
and nuclear physics. This idea has its simplest and most intuitive statement in terms of the
light-cone wavefunction of a high energy hadron or nucleus.  Saturation of quark and antiquark
densities\cite{Gri,Mue} is the statement that the density of quarks per unit area and per unit of
two-dimensional transverse momentum, that is per unit of true transverse phase space, is limited
by a constant times the number of colors so long as the quark momentum is below some momentum 
$Q_s,$ the saturation momentum.  Above $Q_s$ the quark distribution becomes perturbative.  The
result is stated precisely in (29).  While the value of $Q_s$ can depend on the particular hadron
whose wavefunction is being considered and on the longitudinal momentum fraction of the quark, the
statement of saturation, as given in (29), contains no knowledge of QCD dynamics or of the hadron
in question.  The analogous statement of gluon saturation\cite{Gri,Mue,Jal,Kov} is somewhat
different.  Gluon densities can be much larger than quark densities.  The gluon density naturally
has a term ${N_c^2-1\over \alpha N_c}$ as a factor in its ultimate limit and a $\ell n Q_s/\ell^2$
also appears leading to the expression (64) for the gluon density per unit of phase space in the
light-cone wavefunction.  Eqs.(29) and (64) are based on one-loop quark and gluon calculations,
respectively, and for a large nucleus with an additional use of BFKL dynamics in the gluon case. 
However, because (29) and (64) are so directly related, in general, to the scattering of a
quark-antiquark pair and a gluon pair, respectively, on a nucleus or on a high momentum hadron we
believe that these results are quite general except for the possibility of a pure number, not
depending on the hadron in question, as a multiplicative factor on the right-hand sides of (29)
and (64).

In the case of the gluon density in a large nucleus in the saturation region there is a  very nice
model suggested by McLerran and Venugopalan\cite{Ler} where the valence quarks of the nucleons of
the nucleus are treated as the sources of Weizs\"acker-Williams gluons\cite{Jal,Ler,gov} which
make up the small-x gluon distribuiton of the nucleus. This model leads to gluon saturation as
expressed in (6) and (8).  Except for an overall constant factor we believe these are general
results and thus that the Weizs\"acker-Williams model is a good picture of the gluon saturation
regime of the high energy hadron or nucleus.  In the Weizs\"acker-Williams approximation the
saturated gluon distribution is a pure gauge field and this again is believed to be a general
result\cite{Wal}.  The fact that the saturated gluons are pure gauge fields does not mean that they
do not have a direct physical interpretation.  Indeed, our procedure for calculating the
gluon distribution in the nucleus is to note that it is the same as the spectrum of produced
gluons\cite{Kov}, a physical object.  However, the fact the Weizs\"acker-Williams gluons are pure
gauge gluons is what allows an exact calculation of their distribution.

The dynamics that leads to parton saturation is BFKL evolution\cite{Bal,aev} because the
increasing number of gluons that appear in a  hadron's wavefunction due to longitudinal momentum or
x-evolution occupy a common region in transverse phase space and so can naturally lead to the
large values of $A_\mu$ (see (8)) required for saturation.  It is possible that the turnover in
${\partial F_2\over \partial \ell n Q^2}$ observed recently at HERA\cite{wel,Abr} at small\ $x$\ as
$Q^2$ is lowered below $2GeV^2$ may already indicate a saturation of the quark
distribution\cite{ans,man,Gol}.    The fact that diffractive production, $\gamma^*+ P \to x + P,$
indicates an  energy dependence much stronger than that suggested by soft physics\cite{Abr} may
also be an indication that saturation has been reached at HERA\cite{Lev,Eur}.  Deep inelastic
scattering on nuclear targets at HERA should be a very good place to look for saturation. If the
turnover in ${\partial F_2\over \partial \ell n Q^2}$ really is due to saturation one can expect
that the turnover occur at $Q^2 > 5 GeV^2$ in deep inelastic scattering off large nuclei. Another
place where saturation effects may be important is in the very early stages of relativistic heavy
ion collisions.   At RHIC one expects the gluon saturation momentum to be about one GeV with
minijets in that regime contributing most of the freed energy.  At LHC the saturation momentum
should be 2-3 GeV bringing saturation dynamics into play as a major determinant in the very early 
stages, well before equilibration, of heavy ion collisions.  Finally, when potentials as large as
$A \sim 1/g$ are reached one has entered a whole new regime of nonperturbative QCD where, for
example, instanton effects can become important.  Since the large potentials in the wavefunctions
are pure gauge fields perturbation theory remains valid in describing the light-cone
wavefunctions\cite{Wal}.  However, in the very early stages of the central region of a head-on
heavy ion collision these gauge fields are freed, and over the time of the freeing of the gluons
large field strengths, $F_{\mu\nu} \sim 1/g,$ appear whose dynamics should be genuinely
nonperturbative.

In Sec.2,  we indicate how one can determine the quark and gluon distributions in the light-cone
wavefunction by looking at quark and gluon production.  The essence of the argument is that in a
particular light-cone gauge, described in detail in Ref.[4], final state interactions are absent
allowing parton production to be an indicator of the wavefunction of the hadron.  In Sec.2, the
Weizs\"acker-Williams result for the wavefunction is briefly reviewed.

In Sec.3, we calculate the quark momentum distribution in a large nucleus.  This calculation is
equivalent to the one-quark-loop fluctuations in the Weizs\"acker-Williams background field of the
nucleus.  However, our method of doing the calculation, following Ref.[2], emphasizes the
relationship between the quark distribution of the nucleus and the cross section for scattering a
quark-antiquark pair on the nucleus.  Saturation then corresponds to blackness in the scattering
of the quark-antiquark pair on the nucleus.

 In Sec.4, we determine the gluon distribution at the one-loop level. Here the saturated
distribution includes a factor of $\ell n 1/x$ replacing ${1\over \alpha} \ell n Q_s^2/\ell^2$ from
the Weizs\"acker-Williams result.

 In Sec.5, we interpret the one-loop results in terms of unitarity limits and the quantum
mechanical shadow term from a black disc.

 Finally, in Sec.6, we argue that higher corrections, beyond the one-loop correction should simply
replace, up to a constant factor, the $\ell n 1/x$ factor found in the one-loop calculation by
${1\over \alpha} \ell n Q_s^2/\ell^2$ leading again to the semiclassical result.

\section{Determining quark and gluon densities in the light-cone wavefunction}

 The key observation allowing one to measure and calculate quark and gluon distributions in the
light-cone wavefunction is that these distributions are directly related to quark and gluon
production in hard scattering  reactions initiated by currents coupling to quarks and gluons. 
It is a familiar result in deep inelastic electron-proton scattering that the structure function
$F_2$ gives a measure of the quark distributions

$$ F_2(x, Q^2) = \sum_f e_f^2[x q_f(x, Q^2) + x \bar{q}_f(x, Q^2)],$$

\noindent at least in a first-order QCD formalism.  We are now searching a stronger result.  In (1)
the transverse momentum of the struck quark is integrated over the range $0 \leq \ell_\perp^2 \leq
Q^2.$  What we now wish to determine is the quark and gluon distributions in a proton or nucleus
for a definite value of the parton's transverse momentum, $\ell_\perp.$  These unintegrated quark
and gluon distributions are determined by the cross section for producing quarks and gluons at a
definite $\ell_\perp$ in a  deep inelastic reaction.  The idea is to choose the boundary
conditions, the $i\epsilon$'s, of the light-cone gauge so that thre are no final state
interactions\cite{Kov} thus guaranteeing that the struck parton appears in the final state with
unchanged momentum.

\subsection{The quasi-classical approximation}

 This analysis has already been carried out in some detail\cite{Kov} in a quasi-classical
calculation of gluon production off a large nucleus in a deep inelastic reaction initiated by the
``current'' $j= - {1\over 4} F_{\mu\nu}^a F_{\mu\nu}^a$ which couples directly to gluons.  The
result found for the spectrum of produced gluons,   ${dN\over d^2\ell},$ is most easily expressed in
terms of the quantity

\begin{equation}
\tilde{N}(\underline{x}) = \int d^2\ell e^{-i\underline{\ell}\cdot \underline{x}} {dN\over d^2\ell}
\end{equation}

\noindent which is given as\cite{Jal,Kov}

\begin{equation} \tilde{N}(\underline{x}) = \int d^2b {N_c^2-1\over \pi^2\alpha
N_c\underline{x}^2}(1-e^{-\underline{x}^2Q_s^2/4}).
\end{equation}

\noindent In (2) the saturation momentum $Q_s$ is given by

\begin{equation}
Q_s^2 = {8\pi^2\alpha N_c\over N_c^2-1} {\sqrt{R^2-b^2}}\  \rho\  xG(x,\bar{Q}^2)
\end{equation}

\noindent where $\rho$ is the nuclear density and $xG$ is the gluon distribution for a nucleon
with $1/\underline{x}^2 = \bar{Q}^2,$ the scale at which gluons are measured.  $b$  is the impact
parameter of the current-nucleus interaction while  $R$  is the radius of the nucleus.  Eq.2 is
valid in a quasi-classical limit in which $Q^2{\partial\over \partial Q^2} xG(x,Q^2)$ has neither
$Q^2$-dependence nor $\ell n 1/x$ factors at small x.  Two limits of (2) are  noteworthy.  (i)  At
small values of $\underline{x}^2$ 

 \begin{equation}
 \tilde{N}(\underline{x})
 {\longrightarrow_{\underline{x}^2{\rm small}}} x G_A({\rm{x}},\underline{x}^2)
 \end{equation}

\noindent reflecting independent scattering on the various nucleons of the nucleus.  (ii)  For very
large  R

\begin{equation}
\tilde{N}(\underline{x}){\longrightarrow_{ R\    large}}{N_c^2-1\over \pi \alpha N_c}\  {R^2\over
\underline{x}^2}
\end{equation}

\noindent reflecting the saturation of the number of gluons per unit area.  In momentum space,
but neglecting the logarithmic $\underline{x}^2$ dependence of $Q_s^2,$

\begin{equation}
{dN\over d^2bd^2\ell} = {N_c^2-1\over 4\pi^3\alpha N_c} \int_1^\infty {dt\over t}\ 
e^{-t\ell^2/Q_s^2}\  {\longrightarrow_{ \ell^2/Q_s^2 <<1}}\ {N_c^2-1\over 4\pi^3\alpha
N_c} \ell n Q_s^2/\ell^2.
\end{equation}

\noindent This interpretation of saturation is made sharper by noting that

\begin{equation}
-\int d^2b < A_\mu^{i\perp}(\underline{b}) A_\mu^{i\perp
}(\underline{b} +\underline{x}) > = \pi \tilde{N}(\underline{x})
\end{equation}

\noindent so that (6) can be written as\cite{Jal}

\begin{equation}
- < A_\mu^{i\perp}(\underline{b}) A_\mu^{i\perp}(\underline{b}
 + \underline{x}) > = {(N_c^2-1)\over \pi\alpha N_c} {1\over
\underline{x}^2}
\end{equation}

\noindent showing that for gluons of transverse size $\Delta x_\perp$ the maximum value of $\Delta
x_\perp^2 A_\mu^2$ is of order $1/\alpha.$  This is gluon saturation.  The $< >$ in (7) and (8)
indicates averages in the light-cone wavefunction of the  nucleus. We note that satuation is different
than shadowing\cite{Kov} since (4) indicates that there is no shadowing in the quasi-classical
approximation.  Finally, comparing (1) and (7) we note that the number density of gluons in the
light-cone wavefunction as given by (7) is equivalent to the distribution of produced gluons in
deep inelastic scattering.

In Ref.[4], detailed arguments were given that, with a proper choice of light-cone denominators,
final state interactions are  absent in light-cone gauge thus allowing gluon production, at a
given transverse momentum, to directly reflect the transverse and longitudinal momentum
distributions of gluons in the light-cone wavefunction.  In Ref.[4], these arguments were given in
the context of a quasi-classical (Weizs\"acker-Williams) approximation, however the result appears
to be more general  as illustrated in Appendix A of this paper.  In the next two sections of this
paper we calculate first the produced quark distribution and then the produced gluon distribution
in deep inelastic scatterings off a nucleus in the one-loop approximation.  The results can then
be identified with the quark and gluon distributions in a nucleus in the one-loop approximation.

\section{Quark distributions}

In this section, we calculate the produced quark distribution in deep inelastic scattering off
a large nucleus.  Though the calculation is done at the one-loop level we shall argue that the
result is quite general and valid also for scattering off protons at very small values of  x.  The
calculation we are about to perform is not so far different from what has previously been
done\cite{Mue} for the deep inelastic cross section.  The new element which is added here is the
determination of the transverse momentum of the leading quark which then gives the quark
distribution in the light-cone wavefunction.

\subsection{The lowest order}

In order to set normalizations we begin by calculating deep inelastic scattering off a single
nucleon and in the one-loop approximation.  The relevant graphs are shown in Fig.1.  We choose
a frame where

\begin{figure}[htb]
\epsfxsize=3.5in
\epsfbox[0 0 527 161]{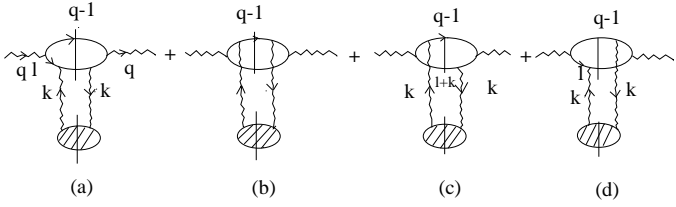}
\caption{Virtual Compton Scattering off a Nucleon in the 
One-quark loop Approximation.}
\label{Fig.1-937}
\end{figure}
a

$$q_\mu = (q_+,q_-,q_\perp) = (-{
q^2\over    2q_-}, q_-,\underline{0})$$
\noindent and
\begin{equation}
(q-\ell)_\mu = ({\ell^2\over 2(1-z)q_-}, (1-z)q_-,-\underline{\ell})
\end{equation}

\noindent and, in addition, we always suppose that $q_- >> Q^2/2q_-$ so that the scattering, say
in a covariant gauge calculation, takes the form of the virtual (transverse) photon breaking up
into a quark-antiquark pair which then scatters off the proton.  The vertical lines running
through the graphs of Fig.1 indicate that the imaging part of the forward Compton amplitude is
taken.  It is straightforward to write

\begin{displaymath}
x q + x\bar{q} = {2\alpha Q^2\over \pi} \sum_\lambda\int {d^2\ell\over 4\pi^2}
dz[z^2+(1-z)^2]
\end{displaymath}

\begin{equation}
\vert \underline{\epsilon}^\lambda\cdot\left({\underline{\ell}\over \ell^2+Q^2z(1-z)} - {(\underline{\ell}+
\underline{k})\over (\underline{\ell}+\underline{k})^2+Q^2z(1-z)}\right
)\vert^2\cdot {d^2k\over [\underline{k}^2]^2} k^2 {\partial
xG(x,k^2)\over \partial k^2}.
\end{equation}

\noindent where

\begin{equation} \epsilon_\mu^\lambda = (\epsilon_+^\lambda, \epsilon_-^\lambda
\underline{\epsilon}^\lambda) = (0,0, \underline{\epsilon}^\lambda)
\end{equation}

\noindent is the polarization of the virtual photon while $xG$ is the gluon distribution of the
target nucleon.  It is not difficult to see that (10) leads to the usual expression for
the quark sea.  In the logarithmic approximation $\underline{k}^2 << \underline{\ell}^2$ the term $\vert\  
\vert^2$ in (10) becomes

\begin{equation}
\sum_\lambda \vert\  \ \vert^2 = {\ell^2\over [\ell^2 + Q^2z(1-z)]^2\ell^2} 
[k^2-{4(\underline{k}\cdot
\underline{\ell})^2 Q^2 z(1-z)\over [\ell^2+Q^2z(1-z)]^2}],
\end{equation}

\noindent which after angular averaging in $\underline{k}$ becomes

\begin{equation}
\sum_\lambda \vert\  \vert^2 = {k^2\over [\ell^2+Q^2z(1-z)]^4}\{(\ell^2)^2+
[Q^2z(1-z)]^2 \}.
\end{equation}

\noindent Again, in the logarithmic aproximation $\underline{\ell}^2 << Q^2$ and $z << 1,$ with
$\underline{\ell}^2$  of the same size as $Q^2z(1-z),$ so that using (13) one finds from (10)

\begin{equation}
x(q(x,Q^2) + \bar{q}(x,Q^2)) = {\alpha\over 3\pi} \int_0^{Q^2} {d\ell^2\over \ell^2} xG(x,\ell^2)
\end{equation}

\noindent which is the correct leading logarithnic form of the DGLAP\cite{Dok,Lip,Par} equation
thus confirming our normalization in (10).

From (10) we an get the differential distribution in transverse momentum as

\begin{displaymath}
{dxq(x,Q^2)\over d^2\ell}  =  {\alpha Q^2\over 2\pi^3}  \sum_\lambda \int_0^1 dz
[z^2+(1-z)^2]
\end{displaymath}

\begin{equation}
\vert \underline{\epsilon}^\lambda\cdot\left({\underline{\ell}\over \ell^2+Q^2z(1-z)} -{\underline{\ell} + \underline{k}\over
(\underline{\ell}+\underline{k})^2+ Q^2z(1-z)}\right)\vert^2{d^2k\over k^2}\ {\partial x G\over
\partial R^2}.
\end{equation}

\noindent It is convenient to go from $\underline{\ell}$ to the conjugate coordinate $\underline{x}$  by using

\begin{equation}
{\underline{\epsilon} \cdot \underline{\ell}\over \ell^2 + Q^2z(1-z)} = \int {d^2x\over 4\pi}
e^{-i\underline{\ell}\cdot\underline{x}}-i\underline{\epsilon}\cdot \bigtriangledown K_0({\sqrt{Q^2x^2z(1-z)}})
\end{equation}

\noindent which gives

\begin{displaymath}\vert\underline{\epsilon}\cdot\bigl({\underline{\ell}\over \ell^2+ Q^2z(1-z)} - {\underline{\ell} +
\underline{k}\over (\underline{\ell} + \underline{k})^2+ Q^2z(1-z)}\bigr)\vert^2\  = \ \int {d^2x_1d^2x_2\over 16\pi^2}
e^{-i\underline{\ell}\cdot(\underline{x}_1-\underline{x}_2)}
\end{displaymath}
\begin{equation} 
(1-e^{-i\underline{k}\cdot\underline{x}_1})(1-e^{i\underline{k}\cdot \underline{x}_2}) \underline{\epsilon}\cdot
\bigtriangledown_{x_1}K_0({\sqrt{Q^2x_1^2z(1-z)}})\underline{\epsilon}\cdot
\bigtriangledown_{x_2}K_0({\sqrt{Q^2x_2^2z(1-z))}}).
\end{equation}

\noindent Thus,

\begin{displaymath}
{dxq\over d^2\ell} = {\alpha Q^2\over 2\pi^3} \int dz {d^2x_1d^2x_2\over 16\pi^2}[(1-e^{-\underline{k}\cdot
\underline{x}_1})+(1-e^{i\underline{k}\cdot \underline{x}_2})-(1-e^{-i\underline{k}\cdot (\underline{x}_1-\underline{x}_2)})]
\end{displaymath}
\begin{equation}
\cdot {d^2k\over
k^2}\ {\partial xG\over \partial k^2}\ [z^2+(1-z)^2]\bigtriangledown_{x_1}K_0({\sqrt{Q^2x_1^2z(1-z)}})\cdot
\bigtriangledown_{x_2}K_0({\sqrt{Q^2x_2^2z(1-z)}}) e^{-i\underline{\ell}\cdot(\underline{x}_1-\underline{x}_2)}.
\end{equation}

\noindent In the logarithnic approximation the exponential factors involving\  $ k$\  in (18) can be
expanded through second order with the resulting integration over\   $ k$\  limited by the
corresponding $\underline{x}-$factor.  Thus,

\begin{equation}
\int (1-e^{-\underline{k}\cdot\underline{x}_1}) {d^2k\over k^2}\ {\partial x G\over \partial k^2}
= {\pi\over 4}\ x_1^2xG(x,x_1^2)
\end{equation}

\noindent where $x_1^2G(x,x_1^2)$ is an abbreviation for $x_1^2 G(x,Q^2=1/x_1^2),$ leading to

\begin{displaymath}
{dxq\over d^2\ell} = {\alpha Q^2\over 128\pi^4} \int d^2x_1d^2x_2dz
[x_1^2xG(x,x_1^2)+x_2^2xG(x,x_2^2)-(x_1-x_2)^2xG(x,(x_1-x_2)^2)].
\end{displaymath}

\begin{equation}
\cdot [z^2+ (1-z)^2]\bigtriangledown_{x_1}K_0\cdot \bigtriangledown_{x_2} K_0
e^{-i\underline{\ell}\cdot(\underline{x}_1-\underline{x}_2)}.
\end{equation}

\subsection{The quark distribution for a large nucleus}

Now consider (20) as the single scattering approximation for quark production on a large nucleus.
Introducing the nuclear density, $\rho,$ and the impact parameter corresponding to the nucleon in
the nucleus, $b,$ we may write (20) as

\begin{displaymath}
{dxq\over d^2\ell} = {Q^2N_c\over 64\pi^6}\ \int d^2bd^2x_1d^2x_2\ [{x_1^2\tilde{v}\over 2\lambda}
{\sqrt{R^2-b^2}}{C_F\over N_c} + {x_2^2\tilde{v}\over 2\lambda}{\sqrt{R^2-b^2}}{C_F\over N_c}
\end{displaymath}
\begin{displaymath} -
{(x_1-x_2)^2\over 2\lambda}{\sqrt{R^2-b^2}}{C_F\over N_c}].
\end{displaymath}

\begin{equation}
e^{-i\underline{\ell}\cdot(\underline{x}_1-\underline{x}_2)}\cdot [z^2+(1-z)^2]dz\bigtriangledown_{x_1}K_0({\sqrt{Q^2x_1^2z(1-z)}})\cdot
\bigtriangledown_{x_2}K_0({\sqrt{Q^2x_2^2z(1-z)}}).
\end{equation}

\noindent In arriving at (21) we have used\cite{Bai}

\begin{equation}
{x_1^2\tilde{v}\over \lambda} = {4\pi^2\alpha N_c\over N_c^2-1} \rho x_1^2 xG(x,x_1^2)
\end{equation}

\noindent where

\begin{equation}
2\int d^2b \rho{\sqrt{R^2-b^2}} = A
\end{equation}

\noindent with \ $A$\  the atomic number of the nucleus.

Now it is straightforward to allow the quark-antiquark pair coming from the virtual photon
to scatter on an arbitrary number of  nucleons in the nucleus.  One simply makes the
replacement\cite{Kov}

\begin{equation}
{x^2\tilde{v}\over 2\lambda}{\sqrt{R^2-b^2}}{C_F\over N_c} \rightarrow 1 -
exp[-{x^2\tilde{v}\over 2\lambda}{\sqrt{R^2-b^2}} {C_F\over N_c}].
\end{equation}

\noindent Introducing, the saturation momentum, $Q_s,$ by

\begin{equation}
Q_s^2 = {2\tilde{v}\over \lambda} {\sqrt{R^2-b^2}}{C_F\over N_c}
\end{equation}

\noindent one finds

\begin{displaymath}
{dxq\over d^2\ell} = {Q^2N_c\over 64\pi^6} \int
d^2bd^2x_1d^2x_2\left(1+e^{-(\underline{x}_1-\underline{x}_2)^2Q_s^2/4}-e^{-x_1^2Q_s^2/4}-e^{-x_2^2Q_s^2/4}\right).
\end{displaymath}
\begin{equation}
e^{-i\underline{\ell}\cdot(\underline{x}_1-\underline{x}_2)}\cdot
dz[z^2+((1-z)^2]\bigtriangledown_{x_1}K_0({\sqrt{Q^2x_1^2z(1-z)}})\cdot\bigtriangledown_{x_2}K_0({\sqrt{Q^2x_2^2z(1-z)}}).
\end{equation}

\noindent We suppose $Q^2 >> Q_s^2.$  Then the dominant contribution to (24) comes from the region
$z << 1.$  It is convenient to define a scaled variable $y=Q^2z$ in terms of which

\begin{displaymath}
{dxq\over d^2bd^2\ell} = {N_c\over 64\pi^6} \int d^2x_1d^2x_2 (1+e^{-(\underline{x}_1-\underline{x}_2)^2 Q_s^2/4}
-e^{-x_1^2 Q_s^2 2/4} -e^{-x_2^2 Q_s^2/4})
e^{-i\underline{\ell}\cdot(\underline{x}_1-\underline{x}_2)}
\end{displaymath}
\begin{equation}
\cdot dy \bigtriangledown_{x_1} K_0\ ({\sqrt{x_1^2y}})\cdot
\bigtriangledown_{x_2}K_0\ ({\sqrt{x_2^2y}}\ ).
\end{equation}

It appears difficult to give a closed form for all the integrals in (27).  However, it is rather
simple to evaluate (27) either when $\ell^2 >> Q_s^2$ or when $\ell^2 << Q_s^2.$  In these cases

\begin{equation}
{dxq\over d^2bd^2\ell} = {N_c\over 6\pi^4}\  {Q_s^2\over \ell^2}\ {\rm for}\  \ell^2 >> Q_s^2
\end{equation}

\noindent and

\begin{equation}
{dxq\over d^2bd^2\ell} = {N_c\over 2\pi^4}\ {\rm for}\  \ell^2 << Q_s^2.
\end{equation}

\noindent In arriving at (29) we have used

\begin{equation}
\int d^2x_1d^2x_2
e^{-i\underline{\ell}\cdot(\underline{x}_1-\underline{x}_2)}
 dy \bigtriangledown_{x_1}K_0({\sqrt{x_1^2y}})\cdot
\bigtriangledown_{x_2}K_0({\sqrt{x_2^2y}} ) = 16\pi^2
\end{equation}

\noindent and

\begin{equation}
\int d^2x_1d^2x_2 e^{-(\underline{x}_1-\underline{x}_2)^2Q_s^2} dy \bigtriangledown_{x_1}K_0({\sqrt{x_1^2y}} )\cdot
\bigtriangledown_{x_2} K_0({\sqrt{x_2^2y}} ) = 16\pi^2
\end{equation}

\noindent while the other two terms in (27) are negligible when $\ell^2 << Q_s^2.$  The integral
(31) coming from the second term on the right-hand side of (27) corresponds to the multiple
scattering of the observed quark by the medium in both the amplitude and in the complex conjugate
amplitude.  The integral (30) coming from the first term on the right-hand side of (27)
is the shadow  term for (31) corresponding to complete blackness of the scattering of the
quark-antiquark pair on the nucleus in the region $\ell^2 << Q_s^2.$

\indent Eq.(29) reflects satuation of the quark density in the nucleus for $\ell^2 << Q_s^2.$  There is, up
to a constant, one quark per unit phase space in the saturation limit.  Unfortunately, we are
unable to give a physical interpretation of the constant appearing on the right-hand side of
(28).  As will be discussed in more detail for the gluon case we believe that, except for the
constant factor, (29) is completely general and that it does not depend on the one-loop
approximation which we have used or on the fact that we have considered a  large nucleus rather than
a hadron with the quark at a very small value of \ $x.$

\section{Gluon distributions}

Now we turn to determining the gluon distribution in our large nucleus.  This has earlier been
done in the quasi-classical approximation.  Here, we do the calculation at the one-loop level, a
calculation from which we shall be able to extract a general result in Sec.6.  The calculation we
are about to perform is related to that done some time ago\cite{Mue}, however, here we focus on
the transverse momentum distribution of the leading gluon, a quantity related to the light-cone
quantized distribution of gluons in the target.

\subsection{The lowest order}

In order to set normalizations we begin with the lowest order calculation.  We use the ``current''

\begin{equation}
j(x) = - {1\over 4} F_{\mu\nu}^i F_{\mu\nu}^i
\end{equation}

\noindent which produces gluons off a nucleon at lowest order from the graphs illustrated in
Fig.2.  We parametrize $q$  and\  $\ell$\  as in (9) while the polarizations of the produced
gluons are written as

\begin{equation}
\epsilon^{\lambda_1}(q-\ell) = (\epsilon_{1+1}^{\lambda_1}, \epsilon_{1-}^{\lambda_1},
\underline{\epsilon}_1^{\lambda_1}) = 
\left(-{\underline{\epsilon}_1^{\lambda_1}\cdot \underline{\ell}\over (q-\ell)_-}, 0,
\underline{\epsilon}_1^{\lambda_1}\right)
\end{equation}

\begin{equation}
\epsilon^{\lambda_2}(\ell + k) = \left({\underline{\epsilon}_2^{\lambda_2}\cdot (\underline{\ell}+\underline{k})\over
\ell_-}, 0, \underline{\epsilon}_2^{\lambda_2}\right)
\end{equation}

\noindent and for shortness of notation we shall often write $\underline{\epsilon}^{\lambda_1} =
\underline{\epsilon}_1$ and $\underline{\epsilon}^{\lambda_2} = \underline{\epsilon}_2.$  In what follows we shall always
work in alogarithmic approximation for longitudinal momentum so that we may assume $\ell_- <<
q_-.$  The sum of the graphs shown in Fig.2 have already been evaluated in Ref.[2] with result, when
$Q^2 >> \ell^2,(\underline{\ell}+\underline{k})^2$ and when $\ell^2, (\underline{\ell}+ \underline{k})^2 << Q^2z$

\begin{figure}[htb]
\epsfbox[0 0 405 136]{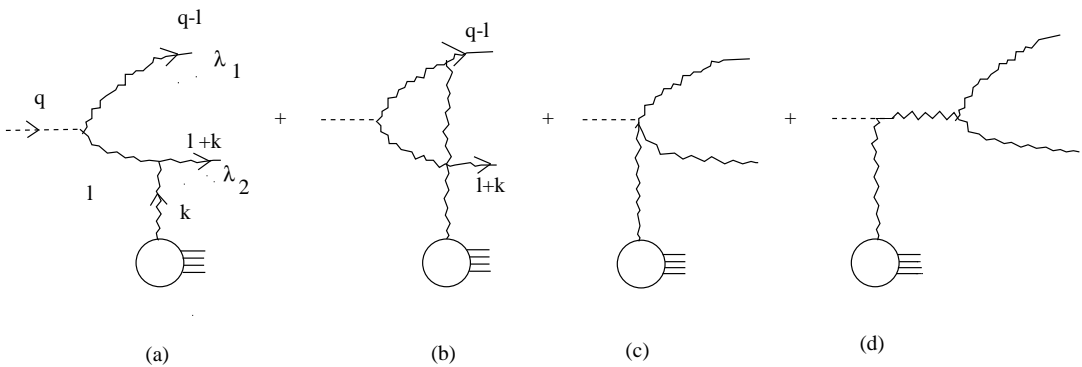}
\caption{Lowest order graphs for gluon pair production off a nucleon.}
\label{Fig.2-937}
\end{figure}

\noindent giving
\begin{equation}
\Gamma_{\lambda_1\lambda_2} = \underline{\epsilon}_1^{\lambda_1}\cdot \underline{\ell}\left[{\underline{\ell}+\underline{k}\over
(\underline{\ell}+\underline{k})^2}\ -\ {\underline{\ell}\over \ell^2}\right] \cdot \underline{\epsilon}_2^{\lambda_2}
\end{equation}

\noindent If the graphs of Fig.2 are interpreted as light-cone perturbation theory graphs, in
contrast to Feynman graphs, and evaluated in $A_- = 0$ gauge then graph\ c\ is zero while the sum
of graphs \ a\ \ and\ b\ is

\begin{equation}
\Gamma_{\lambda_1\lambda_2}^a + \Gamma_{\lambda_1\lambda_2}^b = {\underline{\epsilon}_1\cdot(\ell +
\underline{k})\underline{\epsilon}_2\cdot (\ell + \underline{k})\over (\underline{\ell} + \underline{k})^2} - {\underline{\epsilon}_1\cdot
\underline{\ell} \underline{\epsilon}_2\cdot \underline{\ell}\over \ell^2}.
\end{equation}

\noindent A simple calculation shows that

\begin{equation}
\sum_{\lambda_1\lambda_2}\vert \Gamma_{\lambda_1\lambda_2}\vert^2 = {k^2\over (\underline{\ell}+ \underline{k})^2}
\end{equation}

\noindent while

\begin{equation}
\sum_{\lambda_1\lambda_2}\vert \Gamma_{\lambda_1\lambda_2}^a +
\Gamma_{\lambda_1\lambda_2}^b\vert^2 = 2{\ell^2k^2-(\underline{\ell}\cdot \underline{k})^2\over
\ell^2(\underline{\ell}+\underline{k})^2}
\end{equation}

\noindent so that (37) and (38) are identical when $k^2/\ell^2 << 1$ after an angular average over
directions of $\underline{k}.$  Thus in what follows we consider only graphs (a) + (b), which terms have
an interpretation as the scattering of a color neutral two-gluon system on the target.

\indent For scattering of the current\   $j$\   off a single nucleon we again have four graphs exactly as in
Fig.1 but with $j$ replacing the electromagnetic current $j_\mu$ and with the lines $\ell,
q-\ell, \cdot\cdot\cdot$ now referring to gluons rather than quarks.  Then analogous to (10) one
can write

\begin{displaymath}
xG(xQ^2) = {4\alpha N_c\over \pi} \sum_{\lambda_1,\lambda_2}\int {d^2\ell\over 4\pi^2} {dz\over z}
\end{displaymath}
\begin{equation}
\Vert {\underline{\epsilon}_1\cdot \underline{\ell}\underline{\epsilon}_2\cdot \underline{\ell}\over \ell^2} \ -
{\underline{\epsilon}_1\cdot (\underline{\ell}+\underline{k})\underline{\epsilon}_2\cdot(\underline{\ell}+\underline{k})\over
(\underline{\ell}+\underline{k})^2}\Vert^2{d^2k\over k^2}\ {\partial xG(x,k^2)\over \partial k^2}
\end{equation}

\noindent where, again, we work in an approximation where there are no loops in $xG(x_1k^2)$ so
that $k^2{\partial\over \partial k^2} xG(x,k^2)$ is a constant, both in $\ell n 1/x$ and in
$k^2.$  The $Q^2-$dependence on the left-hand side of (39) comes from a cutoff $\ell^2 < Q^2$
which is understood.  Using (38) in (39) one finds

\begin{equation}
xG(x,Q^2) = \int^{Q^2} {\alpha(\ell^2)N_c\over \pi}\  {d\ell^2\over \underline{\ell}^2}\ xG(x,\ell^2)
\end{equation}

\noindent which is correct in the leading double logarithmic limit, thus checking our normalization
in (39).

Going back to (39) one can write the differential distribution, the unintegrated gluon
distribution, as

\begin{equation}
{dxG\over d^2\ell} = {\alpha N_c\over \pi^3}\sum_{\lambda_1\lambda_2}
\int_{x\ell^2/Q^2}^{\ell^2/Q^2}\ {dz\over z} \vert {\underline{\epsilon}_1\cdot \underline{\ell}\underline{\epsilon}_2\cdot
\underline{\ell}\over \ell^2} 
-{\underline{\epsilon}_1\cdot (\underline{\ell}+\underline{k})\underline{\epsilon}_2\cdot(\underline{\ell}+\underline{k})\over
(\underline{\ell}+\underline{k})^2}\vert^2 {d^2k\over k^2}\ {\partial xG\over \partial k^2}.
\end{equation}

\noindent We note that ${dxG\over d^2\ell}$ is identical to what we alled ${dN\over d^2\ell}$ in
Sec.2, and in Ref.[4].  Writing

\begin{equation}
\underline{\epsilon}\cdot\underline{\epsilon}_2-{2\underline{\epsilon}_1\cdot\underline{\ell}\underline{\epsilon}_2\cdot\underline{\ell}\over
\ell^2} = - \int d^2x\  e^{-i\underline{\ell}\cdot \underline{x}}{1\over \pi x^2}
(\underline{\epsilon}_1\cdot\underline{\epsilon}_2-{2\underline{\epsilon}_1\cdot \underline{x}\underline{\epsilon}_2\cdot\underline{x}\over x^2})
\end{equation}

\noindent one finds

\begin{displaymath}
{dxG\over d^2\ell} = {\alpha N_c\over 2\pi^5}\  \int {dz\over z}\  {d^2x_1d^2x_2\over x_1^2 x_2^2} \
e^{-i\underline{\ell}\cdot(\underline{x}_1-\underline{x})}({2(\underline{x}_i\underline{x}_2)^2\over
x_1^2x_2^2}-1)
 \end{displaymath}
\begin{equation} 
\cdot  (1+e^{-i\underline{k}\cdot(\underline{x}_1-\underline{x}_2)}-e^{-i\underline{k}\cdot\underline{x}_1}-e^{i\underline{k}\cdot\underline{x}_2}){d^2k\over k^2}\ {\partial xG\over \partial k^2}.
\end{equation}

\noindent Duplicating the steps that led from (18) to (26) leads to

\begin{displaymath}
{dxG\over d^2bd^2\ell} = {N_c^2-1\over 8\pi^6} \int {dz\over z}\  {d^2x_1d^2x_2\over x_1^2x_2^2}
e^{i\underline{\ell}\cdot(\underline{x}_1-\underline{x}_2)}(2{(\underline{x}_1\cdot\underline{x}_2)^2\over
x_1^2x_2^2}-1)
\end{displaymath}
\begin{equation}
\left(1+e^{-(\underline{x}_1-\underline{x}_2)^2Q_s^2/4}-e^{-x_1^2Q_s^2/4}-e^{-x_2^2Q_s^2/4}\right)
\end{equation}

\noindent where now

\begin{equation}
Q_s^2 = {2\tilde{v}\over \lambda} {\sqrt{R^2-b^2}}
\end{equation}

\noindent which is identical to (25) except for the absence of the $C_F/N_c$ factor and
${dxG\over d^2\ell} \equiv {dN\over d^2\ell}$ with ${dN\over d^2\ell}$ as defined in Sec.2.

\indent Again, it is easy to evaluate (44) either when $\ell^2 >> Q_s^2$ or when $\ell^2 << Q_s^2.$  Thus,

\begin{equation}
{dxG\over d^2bd^2\ell}={N_c^2-1\over 4\pi^4}\ {Q_s^2\over \ell^2}\ \int\ {dz\over z} =
{N_c^2-1\over 4\pi^4} \ell n\  1/x {Q_s^2\over \ell^2}\ {\rm for}\  \ell^2 >> Q_s^2.
\end{equation}

\noindent When $\ell^2 << Q_s^2$ only the 1 and $e^{-(x_1-x_2)^2Q_s^2}$ terms contribute to (44)
exactly as happened in going from (27) to (29).  Using

\begin{equation}
\int {d^2x_1d^2x_2\over x_1^2 x_2^2}(2{\underline{x}_1\cdot\underline{x}_2)^2\over x_1^2 x_2^2}-1)
e^{i\underline{\ell}\cdot(\underline{x}_1-\underline{x}_2)}=\pi^2
\end{equation}

\noindent and (see Appendix B)

\begin{equation}
\int {d^2x_1d^2x_2\over x_1^2 x_2^2} 
(2{(\underline{x}_1\cdot \underline{x}_2)^2\over x_1^2 x_2^2}-1)
e^{-Q_s^2(\underline{x}_1-\underline{x}_2)^2/4}=\pi^2
\end{equation}

\noindent we arrive at

\begin{equation}
{dxG\over d^2bd^2\ell} = {N_c^2-1\over 4\pi^4} \int_{x Q_s^2RM/Q^2}^{Q_s^2/Q^2} {dz\over z}
\approx {N_c^2-1\over 4\pi^4} \ell n\  1/x
\end{equation}

\noindent where the limits of the $z-$integration are given by assuming the transverse momentum of
each of the gluons approaching the nuclear target is of order $Q_s$ and by requiring the gluonic
system have a coherence length $\geq$ R.  M  is the nucluon mass.  As in the fermion case the
integral (48)  corresponds to absorption of the two-gluon state as it passes over the nucleus 
while the contribution (47) can be viewed as the
quantum mechanical shadow of that absorption and, as usual,  the shadow and the absorption
terms are equal when the target is completely
absorptive (black).  The new element here is the longitudinal momentum integral, the $\ell n\  1/x$ factor in (49).  In
Sec.6, we shall focus on what happens when the calculation is done beyond the one-loop level and
what happens to the $\ell n 1/x$ factor in that case.

\section{Interpreting the one-loop results}

Now, however, let's try to understand the significance of the rather simple results contained in
(46) and (49) as well as in (28) and (29).  Eqs.(28) and (46) of course are straightforward and
represent a process which is hard enough so that only a single nucleon in each nucleus is
effective.  Thus, (28) and (29) give quark and gluon number densities which are, after integrating
over the impact parameter\  $b,$\ just  \ A\ times production off an isolated nucleon.

\indent It is the region where $\underline{\ell}^2 << Q_s^2$ which is more interesting.  Refer for a moment to (27)
where ${dxq\over d^2\ell}$ is given as a sum of four terms on the right-hand side of that
equation. The coordinate $\underline{x}_1$ refers to the transverse position of the observed quark in
the amplitude while $\underline{x}_2$ refers to the same quantity in the complex conjugate amplitude.  The
second term on the right-hand side of (27), the $e^{-(\underline{x}_1-\underline{x}_2)^2 Q_s^2/4}$ term,
corresponds to the S-matrix for the quark-antiquark pair coming from the virtual photon and
interacting with nucleons in the nucleus, both elastic and inelastic interactions, as it passes
over the nucleus. Only the interactions with the observed quark do not cancel between real and
virtual (inelastic and elastic) reactions.  The fact that $(\underline{x}_1-\underline{x}_2)^2 \leq 1/Q_s^2$ shows
that typically the measured quark will have transverse momentum on the order of $Q_s^2.$  Thus the
contribution when $\ell^2 << Q_s^2$ is determined by the probability that a quark which gets many
random ``kicks'' be found with relatively small transverse momentum.  It is natural that $dxq
\propto d^2\ell$ the phase space be available to the quark.  Thus, except for normalization this is
purely a statistical problem for $\ell^2 << Q_s^2.$  The fact that ${dxq\over d^2\ell}$ is
independent of both $Q^2$ and $Q_s^2$ comes from the fact that the quarks which dominate the
process are those having transverse momentum on the order of $Q_s$ \underline{before} the quark-antiquark
pair passes over the nucleus. Quarks having transverse momentum much greater than $Q_s,$ before
reaching the nucleus, are not freed while passing over the nucleus while quarks having transverse
momentum much less than $Q_s,$ before reaching the nucleus, are few in number and can be
neglected.  Since the total number of ``effective'' quarks is proportional to $Q_s^2$ and
distributed according to phase space the functional form, a constant, of (29) follows with
one-half of that constant given by the second term on the right-hand side of (27).

\indent Still in the region $\ell^2 << Q_s^2,$ the first term on the right-hand side of (27) corresponds to
no scattering whatsoever of the quark-antiquark pair by the nucleons of the nucleus. It can be
viewed as the quantum mechanical shadow of the term described just above.  When a quark-antiquark
pair having relative transverse momentum $2\underline{\ell}$ impinges on the nucleus, and if $\ell^2 <<
Q_s^2,$ this pair always interacts with the nucleus with the momentum of the quark and antiquark
getting distorted far from $\underline{\ell}.$  The destruction of this part of the wavefunction is
accompanied by a ``shadow'' term where the quark again has momentum
 $\underline{\ell}.$  This is the first
term on  the right-hand  side of (27).

\indent For the gluon-loop (44),(46) and (49) are direct analogies to (27), (28) and (29) with the new
element being the longitudinal phase space of the gluons.  When the ``current''\   $j$\  breaks up
into a gluon-gluon pair there is a large phase space for one of the gluons (the observed gluon) to
carry almost all the current's longitudinal momentum while the other gluon carries a small amount
which, however, because of the vector nature of the gluon gives a logarithmic integral in the
probability that the current break into a gluon-gluon pair.  It might seem that this gives an
arbitrarily large factor as $\ell n 1/x$ becomes large, but this is not quite so as we shall now
see in the next section.

\indent Finally, we note that when $\ell^2 \leq\   Q_s^2\  {dxG\over d^2\ell}$ and ${dxq\over d^2\ell}$ have an
interpretation as quark and gluon densities in a light-cone wavefunction but they do not have the
interpretation as quark and gluon distributions in terms of an operator product expansion.  This is
most easily seen in the discussion of Sec.2 where ${dN\over d^2\ell} \equiv {dxG\over d^2\ell}$
has significant A-dependence, but where there is no shadowing whatsover and no nontrivial
A-dependence in terms coming from the operator product expansion.

\section{What happens to the $\ell n\  1/x$ factor?}

In this section we shall argue that when higher quantum corrections are included the $\ell n 1/x$
factor in (49) gets modified so as to lead to an expression essentially identical to (6).  To see what happens to the $\ell n\ 1/x$ factor in
(49) we must go beyond the simple small-x dynamics we have used so far in our discussion.  We continue to find it useful to imagine the
scattering of the current \
$j$\ on a large nucleus where we choose an unusual frame in order to explain, heuristically, what is the
essential dynamics.  Thus, choose the kinematics representing the system before the collision
occurs, and illustrated in Fig.3, to be

\begin{figure}[htb]
\epsfbox[0 0 93 124]{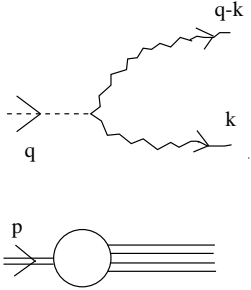}
\caption{Kinematics of the current-nucleus scattering just before the collision.}
\label{Fig.3-937}
\end{figure}

\begin{equation}
q = \left(-{Q^2\over 2q_-}, q_-,0\right)
\end{equation}

\begin{equation}
k = \left({k^2\over 2zq_-}, zq_-, \underline{k}\right)
\end{equation}

\begin{equation}
p=\left(p_+, {m^2\over 2p_+}, 0\right)
\end{equation}

\noindent with \ $p$\ the momentum per nucleon of the nucleus.  With $s = 2q_- p_+$ fixed choose
$q_-$ large enough so that, \underline{for a given z}, $ k_-/k_+$ is significantly larger than one but not
too large.  That is, take
 $q_- = N_0\cdot {\vert\underline{k}\vert\over z}$ with $N_0$ a fixed, and
moderately large, number.  The idea here is to put most of the longitudinal momentum into \ $p$\ 
leaving just enough in \ $q$\  so that the process may be viewed as a two-gluon system, $q-k$ and $k$, colliding with a highly evolved
wavefunction of the nucleus.  By restricting the longitudinal momentum of the left-moving two-gluon system coming initially from  
$j$\ we need not consider further evolution in that system, at least in a leading  logarithmic approximation
in longitudinal momenta. All logarithms except the $dz/z$ integration, which is our focus, are
included in the wavefunction of the nucleus.  Now if  \ $z$\ is decreased one must correspondingly
increase $q_-$ and decrease \ $p_+$\ in order to keep $q_-= N_0 k/z$ and $s=q_-p_+$ fixed.  By
decreasing \ $p_+$\ we limit the range of useful x-evolution in the nucleus and in so doing
decrease the saturation momentum $Q_s.$  But we must guarantee that $k^2 \leq Q_s^2$ in order
that a reaction occur with reasonable probability, and this determines the lower limit of $z$ in
the $dz/z$ integration.  Our task then is to determine the x-dependence of $Q_s^2.$  In the
quasi-classical approximation $Q_s^2,$ given by (3) or (45), has no x-dependence.  However, once
we go beyond the quasi-classical approximation we expect an x-dependence.  Indeed, (3) gives an
x-dependence through $xG(x,Q^2)$ and it is to the determination of the x-dependence of this
quantity that we now turn.

\indent For dynamics we use the fixed coupling BFKL equation which incorporates leading logarithmic
x-evolution, including the small-x approximation to DGLAP evolution. Let 
$xG(\underline{b}, x, Q^2)$ be
the gluon number density for the nucleus at momentum fraction $x,$ at scale $Q,$ and at impact
parameter $b.$  Our normalization is such that

\begin{equation}
\int d^2b\  xG(\underline{b}, x, Q^2) = xG_A(x, Q^2)
\end{equation}

\noindent the normal gluon distribution of the nucleus.  It is $xG(\underline{b}, x, Q^2)$ that we expect
to replace $2{\sqrt{R^2-b^2}}\  \rho x G(x,Q^2)$ in (3) and (45).  Then, in the BFKL approximation

\begin{equation}
xG(\underline{b}, x, Q^2) = \alpha \int N_0(\underline{b}, \lambda, Q_0^2) e^{2{\alpha N_c\over
\pi}\chi(\lambda)Y+\lambda\ell n\ Q^2/Q_0^2}\ {d\lambda\over 2\pi\ i}
\end{equation}

\noindent where we expect $N_0$ to be slowly varying in $\lambda.$  In Eq.(54)

\begin{equation}
\chi(\lambda) = \psi(1) -{1\over 2} \psi (\lambda) - {1\over 2} \psi (1-\lambda)
\end{equation}

\noindent where the $\lambda$-integration goes along the imaginary axis, while $Y=\ell\  n 1/x.$  The
saddle point of the $\lambda$-integration is determined by

\begin{equation}
\chi^\prime(\lambda_0) = - {\ell n\ Q^2/Q_0^2\over {2\alpha N_c\over \pi}\ Y}
\end{equation}

\noindent giving

\begin{equation}
xG(\underline{b}, x, Q^2) = {\alpha N_0(\underline{b},\lambda_0,Q_0^2)\over{\sqrt{4\alpha N_c\ 
\chi^{\prime\prime}
(\lambda_0)Y}}}\  \bigl({Q^2\over Q_0^2}\bigr)^{\lambda_0} e^{{2\alpha N_c\over
\pi}\chi(\lambda_0)Y}.
\end{equation}

\noindent We expect (57) to be valid so long as $Q^2 >> Q_s^2$ with $Q_s^2$ being determined by

\begin{equation}
xG(\underline{b}, x, Q_s^2) = {c\over \alpha} \left({Q_s^2\over Q_0^2}\right)
\end{equation}

\noindent where $xG(\underline{b}, x, Q_s^2),$ approached from the perturbative regime $Q^2 > Q_s^2$ 
agrees with (49) when $\ell^2 \to Q_s^2$ from the lower momentum side, $\ell^2 < Q_s^2.$  We allow\ c\ 
to have weak (logarithnmic)  \ x\ and $Q_s^2$ dependences.  Using (57) and (58) one finds

\begin{equation}
(1-\lambda_0) \ell n\ Q_s^2/Q_0^2 = \ell n \{{N_0\alpha^2\over {c\sqrt{4\alpha N_c\chi^{\prime\prime}Y}}}\}  + {2\alpha
N_c\chi(\lambda_0)\over \pi}Y.
\end{equation}

\noindent Using (56) in (58) gives

\begin{equation}
[1-\lambda_0 + {\chi(\lambda_0)\over \chi^\prime(\lambda_0)}] \ell n\ Q_s^2/Q_0  = \ell n\{{N_0\alpha^2\over {c \sqrt{4\alpha
N_c\chi^{\prime\prime}Y}}}\}.
\end{equation}

\noindent The right-hand side of (60) is slowly varying in \ $Y$\ and in $\lambda_0$.  Thus, for very large\ $Y,$\ leading to very large
$Q_s^2,
\lambda_0$ is determined by

\begin{equation}
1 - \lambda_0 + {\chi(\lambda_0)\over \chi^\prime(\lambda_0)} = 0
\end{equation}

\noindent a value of $\lambda_0$ which is not too far from $\lambda = 1/x$ so that $\ell n\ 1/x$ evolution dominates $Q^2-$evolution
indicating that our appoach to the problem should be reasonable.

\indent Turning to (59) and using the fact that the first term on the right-hand side of that equation is slowly varying in $Y$ one can
determine that

\begin{equation}
{1\over Q_s^2} {d Q_s^2(Y)\over dY} = {2\alpha N_c\chi(\lambda_0)\over \pi(1-\lambda_0)}
\end{equation}

\noindent giving the dependence of the saturation momentum on $Y.$

\indent Now we are in a position to answer the question of what happens to the $\ell n\ 1/x$ factor in (49) when higher quantum corrections
are included.  Let $Y=\ell\ n\ s/Q^2$ and let $Y(\ell)$ be that rapidity such $Q_s^2(Y(\ell)) = \ell^2.$  Then the $z-$integral in (49) becomes

\begin{equation}
\int_{x_0\ \ell^2/Q^2}^{\ell^2/Q^2}\ {dz\over z} = \ell n\ 1/x_0 = \ell n {Q_s^2(Y)\over Q_s^2(Y(\ell))}
\end{equation}

\noindent where $\ell n\ 1/x_0 = Y-Y(\ell)$ so that, using (63) one gets (49) to become

\begin{equation}
{dxG\over d^2bd^2\ell} = {N_c^2-1\over 4\pi^3\alpha N_c}\ {1-\lambda_0\over 2\chi(\lambda_0)}\ \ell n \ {Q_s^2(Y)\over \ell^2}
\end{equation}

\noindent which, apart from the ${1-\lambda_0\over 2\chi}$ factor, is identical to (6).  We feel that the essential factors in (64), the
${N_c^2-1\over \alpha N_c}$ and the $\ell n\ Q_s^2/\ell^2$ factors, are general results in QCD for the light-cone wavefunction.  The overall
constant in (64) we do not trust.

\indent It is perhaps useful to consider carefully why we claim that (49) is general, except for the issue of the $\ell n\ 1/x$ factor which
we have discussed in some detail in this section.  To that end turn to (44).  Except for the four terms in parentheses at the end of the
right-hand side of (44) all the other factors reflect the current $j$ breaking up into a gluon-gluon pair along with the Fourier transform
going from transverse coordinate to transverse momentum space.  Thus all the dynamics of the target is in the last factor. Of the four terms
constituting the last factor the first, the 1, corresponds to no interactions of the gluon pair with the target and this term is universal
and independent of the nature of the target.  The third and fourth term correspond to possible interactions only in the amplitude and complex
conjugate amplitudes, respectively.  These two terms are functions only of $x_1^2$ and $x_2^2,$ respectively and so must generally integrate
to zero when $\ell^2 << Q_s^2.$  The second term corresponding to $S(\underline{x}_1) S^*(\underline{x}_2),$ the product of the $S-$matrices in the amplitude
and complex conjugate amplitude, need not in general take the form given in (44).  This term in general must be 1 when $\underline{x}_1 = \underline{x}_2$ and
it should be small when $(\underline{x}_1-\underline{x}_2)^2 Q_s^2 >> 1$, but we have no argument as to the exponential form of (44) being exact.   It is this
specific form which gives (47) and (48) the same value.  Physically, the exponential form in (44)  came from the many independent scatterers in
the nucleus.  While this is also natural in our more general circumstance we do not know how to prove it.  In any case, the expectation that
the term in question go to zero when
$(\underline{x}_1-\underline{x}_2)^2 Q_s^2 >> 1$ immediately gives a contribution which is
$\underline{\ell}-$independent when
$\ell^2 << Q^2$ since the $\underline{\ell}-$dependence only comes in through $e^{-i\underline{\ell}\cdot (\underline{x}_1-\underline{x}_2)}.$  The real issue then is what
constant replaces the $\pi^2$ on the right-hand side of (48).  We have written (64) as if that constant remains $\pi^2$ while it may possibly
be some other pure number. Thus the form given in (64) we feel must be
true, but there may be an additional constant multiplying the right-hand side of that equation.
\vskip20pt

\noindent{\bf Appendix  A}

\indent In this appendix we illustrate how, with a proper choice of $i\epsilon$'s in light-cone denominators, final state interactons can be
suppressed.  The example given here is similar to that given some time ago
\cite{Qiu} although now we have a better physical interpretation of
what is happening.

\indent Consider the graph shown in Fig.4 where there is a final state interaction of the gluon (k) with the struck quark. In light-cone gauge
only the term ${i\over k^2+i\epsilon}\ {\eta_\alpha k_\beta^\perp\over k_+}$  is important in the gluon propagator.  $p$\ has a large\ +\
component of the momentum, and we suppose \ $q$\ has only\ +\ and --- components with $2q_+q_-  =  - Q^2.$  We may assume that all lines in the
upper blob have \ +  components of their momentum greater or equal to ($\ell-k-q)_+=x p_+ {\rm while} \vert(\ell - q)_- \vert << q_-.$  Thus
$(\ell-k-q)^2 \approx - (\underline{\ell}-\underline{k})^2$ and $k^2 \approx - \underline{k}^2$ so that one need only consider $(\ell-k)^2 + i \epsilon$ and $k_+$ as
denominators possibly trapping the $k_+ -$contour and thus limiting $k_+$ to small values.  It is only in case the $k_+-$contour is trapped at
a value where $\vert k_+\vert \leq {k_\perp^2\over \ell_-}$ that final state interactions are important.  Now $(k-\ell)^2 + i\epsilon \approx -
2\ell_-((k-\ell)_+ + ({(\underline{k}-\underline{\ell})^2\over 2\ell_-}-i\epsilon).$  Thus, if the light-cone denominator is taken to be
$[k_+-i\epsilon]^{-1}$  there is no trapping of the $k_+$ contour while any other choice leads to trapping.  This choice of $i\epsilon$
corresponds to the gauge potential extending to large negative $x_ - -$values, but not to large positive $x_--$ values\cite{Kov,gov}, thus
naturally avoiding final state interactions.  It is straightforward to include additional gluons  connecting to the struck quark.  What is not
so clear is whether or not there is a consistent definition of light-cone denominators which renders higher loops finite and at the same time
eliminates the gauge field for large positive values of $x_-.$  This is an important technical problem yet to be resolved.
\vskip 20pt

\begin{figure}[htb]
\epsfbox[0 0 140 118]{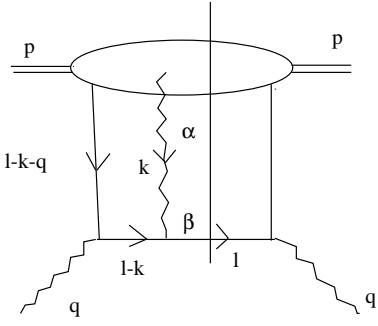}
\caption{Potential final-state interaction which is absent with appropriate choice of boundary conditions for the 
light-cone gauge propagator.}
\label{Fig.4-937}
\end{figure}

\noindent{\bf Appendix B}

In this appendix we outline how the integral in (48) can be evaluated.

$$I = \int {d^2x_1d^2x_2\over  x_1^2x_2^2} (2{(\underline{x}_1\cdot \underline{x}_2)^2\over x_1^2x_2^2} -1) e^{-(\underline{x}_1-\underline{x}_2)^2 Q_s^2/4},\eqno(B1)$$

\noindent then

$$I = {1\over 2} \int {d^2x_1d^2x_2\over  (x_1^2)^2 (x_2^2)^2}[16{\partial^2\over \partial(Q_s^2)^2} + 8(x_1^2+x_2^2){\partial\over \partial
Q_s^2} + (x_1^2)^2 + (x_2^2)^2]e^{-(\underline{x}_1-\underline{x}_2)^2Q_s^2/4}\eqno(B2)$$

\noindent Using

$${1\over 2\pi} \int_0^{2\pi} d\phi e^{{1\over 2} x_1x_2 cos \phi Q_s^2} = I_0({1\over 2}Q^2x_1x_2).\eqno(B3)$$

\noindent One finds

$$I = \pi^2 \int_0^\infty {dx_1^2 dx_2^2\over x_1^2 x_2^2} I_2({1\over 2} Q_s^2 x_1x_2) e^{-(x_1^2+x_2^2)Q_s^2/4}.\eqno(B4)$$

\noindent Now (See formula 19 on page 197, Ref.\cite{Erd}.

$$\int_0^\infty {dx_2^2\over x_2^2} I_2 ({1\over 2} Q_s^2 x_1 x_2) e^{-x_2^2Q_s^2/4} = {1\over Q_s^2 x_1^2}
e^{x_1^2Q_s^2/4} \gamma(2,Q_s^2x_1^2/4)$$

\noindent where $\gamma(\alpha, x)$ is the incomplete $\gamma-$function.  We find

$$I=\pi^2 \int_0^\infty {dz\over z^2} \gamma(2,z) = \pi^2.\eqno(B5)$$

\end{document}